\newif\iflinenumbers
\linenumbersfalse %

\documentclass[10pt,oneside,notitlepage,abstracton,a4paper]{article}

\usepackage{epsfig,scrpage2}
\usepackage{amsmath}
\usepackage{makecell}
\usepackage{graphicx}
\usepackage{booktabs}
\usepackage{arydshln}
\usepackage{setspace} %
\usepackage[utf8]{inputenc}
\usepackage[T1]{fontenc}

  \date{\normalsize \today}

\usepackage{chngcntr}

\usepackage[bottom]{footmisc}

\usepackage{titlesec}
\titleformat*{\section}{\large\bfseries}
\titleformat*{\subsection}{\normalsize\bfseries}

\iflinenumbers
  \usepackage{lineno}
  \linenumbers
\fi

\usepackage[affil-it]{authblk}

\usepackage{etoolbox} %
\makeatletter %
\patchcmd{\@maketitle}{\LARGE \@title}{\fontsize{16}{19.2}\selectfont\@title}{}{}
\makeatother
\usepackage[affil-it]{authblk}

\usepackage{geometry}
\usepackage{xcolor}
\definecolor{DESYcyan}{RGB}{0,166,235}
\definecolor{DESYorange}{RGB}{242,142,0}
\definecolor{DESYgray}{RGB}{119,119,119}
\definecolor{bjetpurple}{RGB}{117,112,179}

\usepackage[margin=8mm,font=small,labelfont=bf,format=plain]{caption}
\usepackage[margin=8mm,font=small,labelfont=bf,format=plain]{subcaption}

\usepackage{multirow} %

\newcommand{\subfigref}[1]{({\protect\subref{#1}})}

\usepackage{slashed}

\newcount\colveccount
\newcommand*\colvec[1]{
        \global\colveccount#1
        \begin{pmatrix}
        \colvecnext
}
\def\colvecnext#1{
        #1
        \global\advance\colveccount-1
        \ifnum\colveccount>0
                \\
                \expandafter\colvecnext
        \else
                \end{pmatrix}
        \fi
}

\usepackage{lmodern}
\usepackage{listings}
\usepackage[english]{babel}%
\usepackage{csquotes}%
\usepackage[style=numeric-comp,sorting=none]{biblatex}
\newcommand{\customprintbibliography}{%
  \printbibliography[heading=none]
  \newpage
}

\usepackage{amssymb}

\newcommand{\Pol}{\mathcal{P}} %

\newcommand{\eP}{e^{+}}
\newcommand{\eM}{e^{-}}
\newcommand{\muP}{\mu^{+}}
\newcommand{\muM}{\mu^{-}}
\newcommand{\lP}{l^{+}}
\newcommand{\lM}{l^{-}}

\newcommand{\nubar}{\bar{\nu}}

\newcommand{\qbar}{\bar{q}}

\newcommand{\WP}{W^{+}}
\newcommand{\WM}{W^{-}}

\title{Precision measurements of Triple Gauge Couplings at future electron-positron colliders}
\author[1,2]{Jakob Beyer\thanks{\textit{\footnotesize{Talk presented at the International Workshop on Future Linear Colliders (LCWS2019),}\\ \footnotesize{Sendai, Japan, 28 October - 1 November, 2019.}\\ \footnotesize{C19-10-28.}}}}
\author[1]{Robert Karl}
\author[1]{Jenny List}
\affil[1]{Deutsches Elektronen-Synchroton (DESY), Hamburg, Germany}
\affil[2]{Universit\"at Hamburg, Hamburg, Germany}
\date{}

\begin{document}

\maketitle

\begin{abstract}
  A precise knowledge of charged Triple Gauge Couplings (cTGCs) is important for the determination of Higgs couplings and for constraining physics beyond the Standard Model.
  Future high-energy $\eP\eM$ colliders could have a significantly improved sensitivity to anomalous cTGCs.
  The fit framework presented here extracts cTGCs in parallel with chiral cross sections and beam polarisation parameters.
  It demonstrates that cTGCs can be measured with precision in the $10^{-3}-10^{-4}$ range.
  A strong dependence of the cTGC sensitivity on the available luminosities and polarisations is observed.
\end{abstract}

\clearpage

\section{Introduction}

The Standard Model (SM) gauge group allows and requires the coupling of three electroweak gauge bosons.
Only interactions including two oppositely charged bosons are allowed in the SM ($\WP\WM Z$,$\WP\WM\gamma$).
The strength of this SM interaction is determined by the gauge structure and depends only on the weak mixing angle $\theta_{W}$ and coupling constant $g$.

Additional variations in the charged triple gauge coupling (cTGC) may occur in the presence of physics beyond the SM (BSM).
Such BSM effects have a strong interplay with precision Higgs coupling measurements~\cite{Bambade:2019fyw}.
These effects must be quantified in a theoretical framework that can describe deviations from the SM prediction.
SM effective field theory provides such a framework and can describe anomalous cTGCs.

Extracting cTGC values from collision data requires a fit to observed differential distributions.
The measurement of the cTGCs uses the same processes which are used for the measurement of the beam polarisations.
A fit of cTGCs must therefore include the beam polarisation parameters.
Other electroweak parameters can interfere with the measurement of cTGC parameters and must be considered as well.

Four proposals of future high-energy $\eP\eM$ collider which could perform such a measurement are currently under discussion.
Two of them are linear machines - ILC~\cite{Aihara:2019gcq} and CLIC~\cite{Roloff:2018dqu} - and two are circular - FCC-ee~\cite{Benedikt:2651299} and CEPC~\cite{EPPSUCEPC}.
The machine parameters most crucial to the physics performance are its collision energies, the luminosities at theses energies and the polarisation of the beams.
These parameters can vary by orders of magnitude between the four proposals due to the different challenges of linear and circular machines.
Collision energies range from runs at the $Z$ pole to the multi-TeV scale.
Achievable luminosities are dependent on machine choice and energy.
At the scale of a Higgs factory ($\sim 250\,$GeV) typical luminosities are around $5\,$ab$^{-1}$ for circular machines and around $1-2$ab$^{-1}$ for linear ones.
The current proposals include options with each, one or no beam polarised. 

This study aims to determine how these top-level collider parameters influence electroweak precision measurements.

\section{Fit of electroweak and polarisation parameters}

A $\chi^2$ minimization is used to vary the parameters and thereby the predictions of differential distributions in order fit them to measurements of those distributions.
Bin-by-bin correlations are not implemented in the current framework and are planned to be implemented in the future.

The fit can vary the three cTGCs ($g_1^Z$, $\kappa_{\gamma}$, $\lambda_{\gamma}$) of the LEP parameterisation~\cite{Hagiwara:1986vm} as well as the polarisation and the chiral cross sections of processes.
A non-zero beam polarisation is assumed to not change its amplitude when its helicity direction is reversed.
This results in a maximum of two beam polarisation parameters - one for the $\eM$ polarisation and one for the $\eP$ polarisation.
Two parameters per scattering process describe possible changes in the allowed chiral cross section.
The first is an asymmetry between two chiral cross sections and the second is the sum of all four chiral cross sections.

In total, the fit has to extract three cTGS parameters, between zero and two polarisation parameters, and two additional parameters for every considered process.

\subsection{Input channels}

  \begin{table} \newcommand{\linedist}{0.2em}
    \centering
    \caption{%
      List of processes used in the fitting framework, including the considered values of chiral chross-sections and the differential distributions with corresponding binning.
      A starred observable ($^*$) is extracted in the rest frame of the corresponding $W$ boson.
      ($q=u,d,s,c,b$)
      Cross sections where calculated with \texttt{WHIZARD}~\cite{Kilian:2007gr} and are supplied by the ILD generator group~\cite{webiste:genlog250}.
    }
    \label{tab:InputProcesses}
    \begin{tabular}{|l|l|l|l|l|l l l|}
      \hline
      Process & $\sigma_{LR}$ & $\sigma_{RL}$ & $\sigma_{LL}$ & $\sigma_{RR}$ & \multicolumn{3}{l|}{Diff. observ.} \\
      ($\eP\eM \rightarrow X$) & $[\text{fb}^{-1}]$ & $[\text{fb}^{-1}]$ & $[\text{fb}^{-1}]$ & $[\text{fb}^{-1}]$ & \multicolumn{3}{l|}{\# Bins} \\[\linedist]\hline
      $\WP\WM \rightarrow \mu\nu q\qbar'$ & 9390 & 86.4 & 0 & 0 & $\cos\left(\theta_{\WM}\right)$, & $\cos\left(\theta_{\mu}^*\right)$, & $\phi_{l}^*$ \\[\linedist]
      & & & & & $20\,\otimes$ & $10\,\otimes$ & $10$ \\[\linedist]\hline
      $\WP\eM\nubar \rightarrow q\qbar'\eM\nubar$ & 5000 & 42.8 & 0 & 119 & $\cos\left(\theta_{\WP}\right)$, & $\cos\left(\theta_{\eM}^*\right)$, & $m_{\eM \nubar}$ \\[\linedist]
      & & & & & $20\,\otimes$ & $10\,\otimes$ & $20$ \\[\linedist]\hline
      $\WM\eP\nu \rightarrow q\qbar'\eP\nu$ & 500 & 42.9 & 120 & 0 & $\cos\left(\theta_{\WM}\right)$, & $\cos\left(\theta_{\eP}^*\right)$, & $m_{\eP \nubar}$ \\[\linedist]
      & & & & & $20\,\otimes$ & $10\,\otimes$ & $20$ \\[\linedist]\hline
      $ZZ \rightarrow q\qbar\muP\muM$ & 356 & 178 & 0 & 0 & $\theta_{Z}^{l}$, & $\theta_{\muM}^*$, & $\phi_{\muM}^*$ \\[\linedist]
      & & & & & $20\,\otimes$ & $10\,\otimes$ & $10$ \\[\linedist]\hline
      $q\qbar$ & 12900 & 71300 & 0 & 0 & $\theta_{q}$ & & \\[\linedist]
      & & & & & $20$ & &  \\[\linedist]\hline
      $\lP\lM (l=\mu/\tau)$ & 21200 & 16500 & 0 & 0 & $\theta_{\lM}$ & & \\[\linedist]
      & & & & & $20$ & &  \\[\linedist]\hline
    \end{tabular}
  \end{table}
  
  A large number of processes and differential cross section bins is required to constrain all parameters of the fit.
  Here, six processes with two or four fermions in the final state are used (tab.~\ref{tab:InputProcesses}).
  
  Generator level datasets are used to calculate the differential distributions for processes with two fermions in the final state.
  The events are generated with \texttt{WHIZARD}~\cite{Kilian:2007gr} and include effects from initial state radiation (ISR) and from the energy spectra of the incoming beams.
  Differential distributions for processes with four fermions in the final are obtained from matrix element calculation with \texttt{O'Mega}~\cite{Moretti:2001zz} and do not consider ISR or beam spectra.
  The distribution from matrix element calculation have previously been compared to corresponding ones which were extracted from Monte Carlo events including ISR and beam spectra and the difference was found to be small~\cite{Karl:424633}.
  
  Four-fermion final states are affected by changes in the cTGCs.
  A three-parameter second order polynomial is used to parameterise this effect for each bin of the distributions.
  The coefficients of the polynomial are calculated from values of the differential cross section at different cTGC values~\cite{Karl:424633}.
  For different values of polarisation and chiral cross sections the differential distributions are re-normalised accordingly.
  
  \begin{figure}
    \centering
    \begin{subfigure}[t]{0.5\textwidth}
      \centering
      \includegraphics[width=\textwidth]{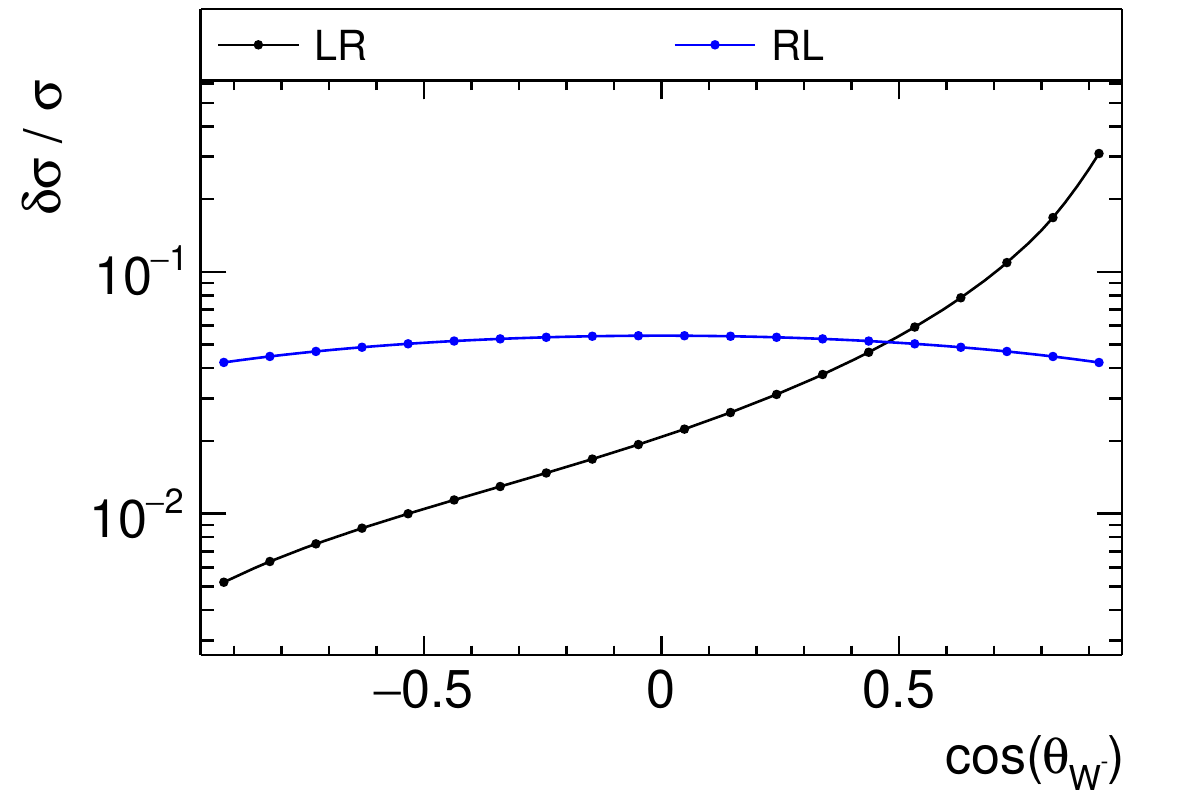}
      \caption{}
      \label{SUBFIG:DiffDistrExample_cosThetaW}
    \end{subfigure}%
    \begin{subfigure}[t]{0.5\textwidth}
      \centering
      \includegraphics[width=\textwidth]{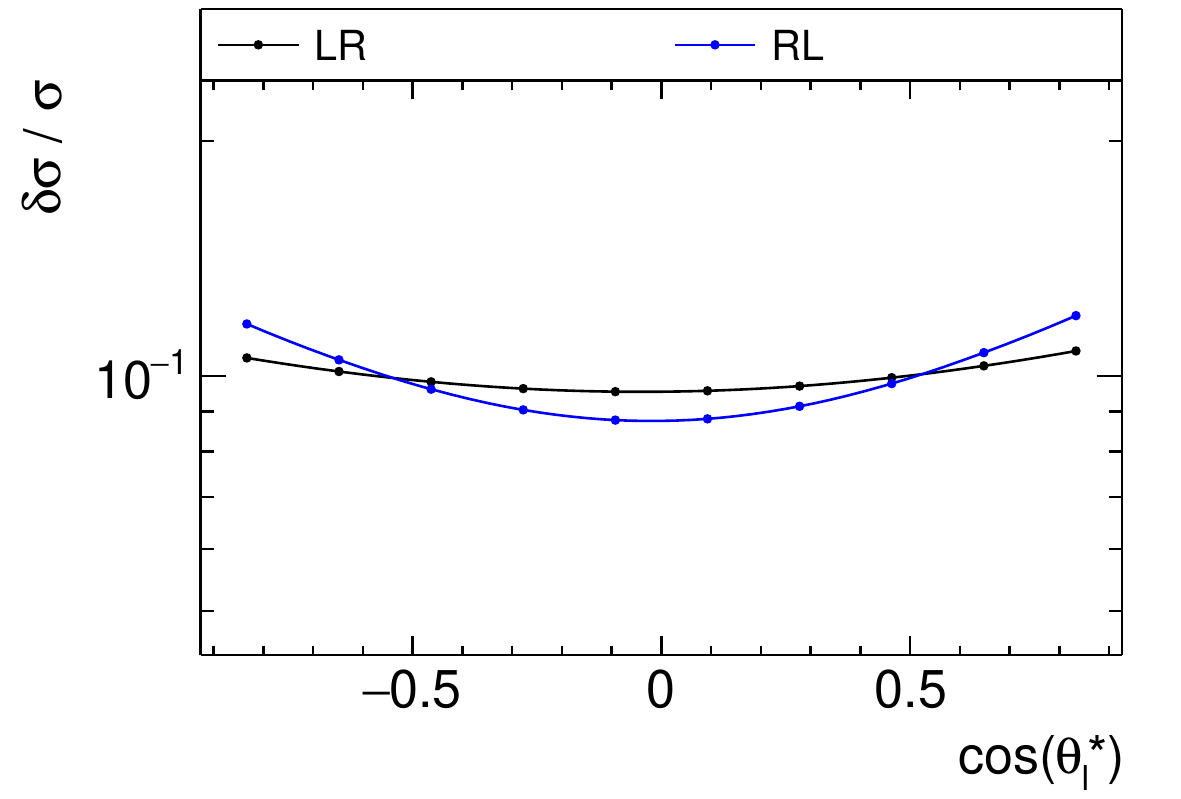}
      \caption{}
      \label{SUBFIG:DiffDistrExample_cosThetal}
    \end{subfigure}
    \caption{%
    Normalised differential cross section distributions of $\eM\eP \rightarrow \WM\WP \rightarrow l{\nu}q\qbar'$ collision at $250\,$GeV for different chiralities of the incoming $\eM\eP$ (e.g. ``LR'' for $\eM_L\eP_R$)~\cite{Karl:424633}.
    \subfigref{SUBFIG:DiffDistrExample_cosThetaW} Cosine of the polar angle of the $\WM$ boson.
    \subfigref{SUBFIG:DiffDistrExample_cosThetal} Cosine of the charged lepton polar angle in the corresponding $W$ rest frame.
    }
    \label{fig:DiffDistrExample}
  \end{figure}
  
  The observables for each process are chosen to fully describe the process kinematics assuming fixed $\sqrt{s}$ and no ISR~\cite{Karl:424633}.
  For two-fermion final states only the polar angle $\theta$ of the particle in the final state is used (e.g. $\theta_q$ for $q\qbar$).
  The azimuthal angle distribution is expected to be flat and does not carry additional information.
  In the four-fermion final states the angles in the center-of-mass frame of the hard $\eP\eM$ collision and the boson rest frame are used.
  The azimuthal angle of the bosons is ignored due to its flat distribution and only the polar angle of the $\WM$ and the leptonically decaying $Z$ boson is used.
  Both polar angles of the (negatively) charged lepton in the leptonically decaying $W$ ($Z$) rest frame are used (e.g. fig.~\ref{fig:DiffDistrExample}).
  Angles in the rest frame of the hadronically decaying $W$ are not used due to experimental ambiguity.
  
  Processes with taus or more than one neutrino in the final state are not considered.
  The $\WP\WM \rightarrow \mu\nu q\qbar'$ channel is not separated by the muon charge.
  A separation of muon charges does not pose an experimental challenge and is only neglected here for techinal simplification.
  This however results in an artificial symmetrisation of the corresponding angular distributions.
  Parameters which are extracted from these differential distributions will therefore have overestimated uncertainties.
  
\subsection{Toy measurements}

  The fit parameter uncertainties achieved within this framework are studied for various collider setups.
  Such a collider setup is a set of energy, luminosity, available polarisations and luminosity sharing between polarisation settings.
  The performance is tested using a fit to toy measurements.
  
  A toy measurement is defined as the set of differential distributions based on their SM expectation for which each bin content gets fluctuated with a Poisson distribution for its statistical uncertainty.
  Migrations between bins are not considered.

  A selection efficiency of $\epsilon = 60\%$ and a purity of $\pi = 80\%$ is assigned to each bin of each distribution. 
  These values are motivated by a full analysis of $l{\nu}qq$ events at $\sqrt{s}=500\,$GeV which included ILD full detector simulation and reconstruction~\cite{Marchesini:2011aka}.
  This models the effect of an analysis on the number of events.
  An analysis of fully simulated datasets of each channel may influence the shape and normalisation of each distribution more subtly.
  This is not yet considered.
  
  Only the beam polarisations are directly considered as systematic uncertainties.
  Their uncertainties taken into account by letting them vary in the fit.
  In addition, the chiral cross section parameters may be interpreted as a reflection of theory uncertainties.
  Such systematic uncertainties from the theoretical cross section calculations would be absorbed into an extraction of anomalous chiral cross section values.
  Beyond that, no other systematic uncertainties - such as those on luminosity or detection efficiency - are included.

\section{Results}

  \begin{table} \newcommand{\linedist}{0.2em}
    \centering
    \caption{%
      Luminosity sharing in the studied setups.
    }
    \label{tab:lumisharing}
    \begin{tabular}{|l|l l l l l l l|}
      \cline{2-8}
      \multicolumn{1}{l|}{} & \multicolumn{7}{l|}{Luminosity sharing ($\text{sgn}(P_{\eM})$,$\text{sgn}(P_{\eP})$) [\%]}\\ \hline
      Polarisation setup & (--,+) & (+,--) & (--,--) & (+,+) & (--,0) & (+,0) & (0,0) \\ \hline
      both polarised & 45 & 45 & 5 & 5 & - & - & - \\
      only $\eM$ polarised & - & - & - & - & 80 & 20 & - \\
      both unpolarised & - & - & - & - & - & - & 100 \\ \hline
    \end{tabular}
  \end{table}

\begin{figure}
  \centering
  \begin{subfigure}[t]{0.5\textwidth}
    \centering
    \includegraphics[width=\textwidth]{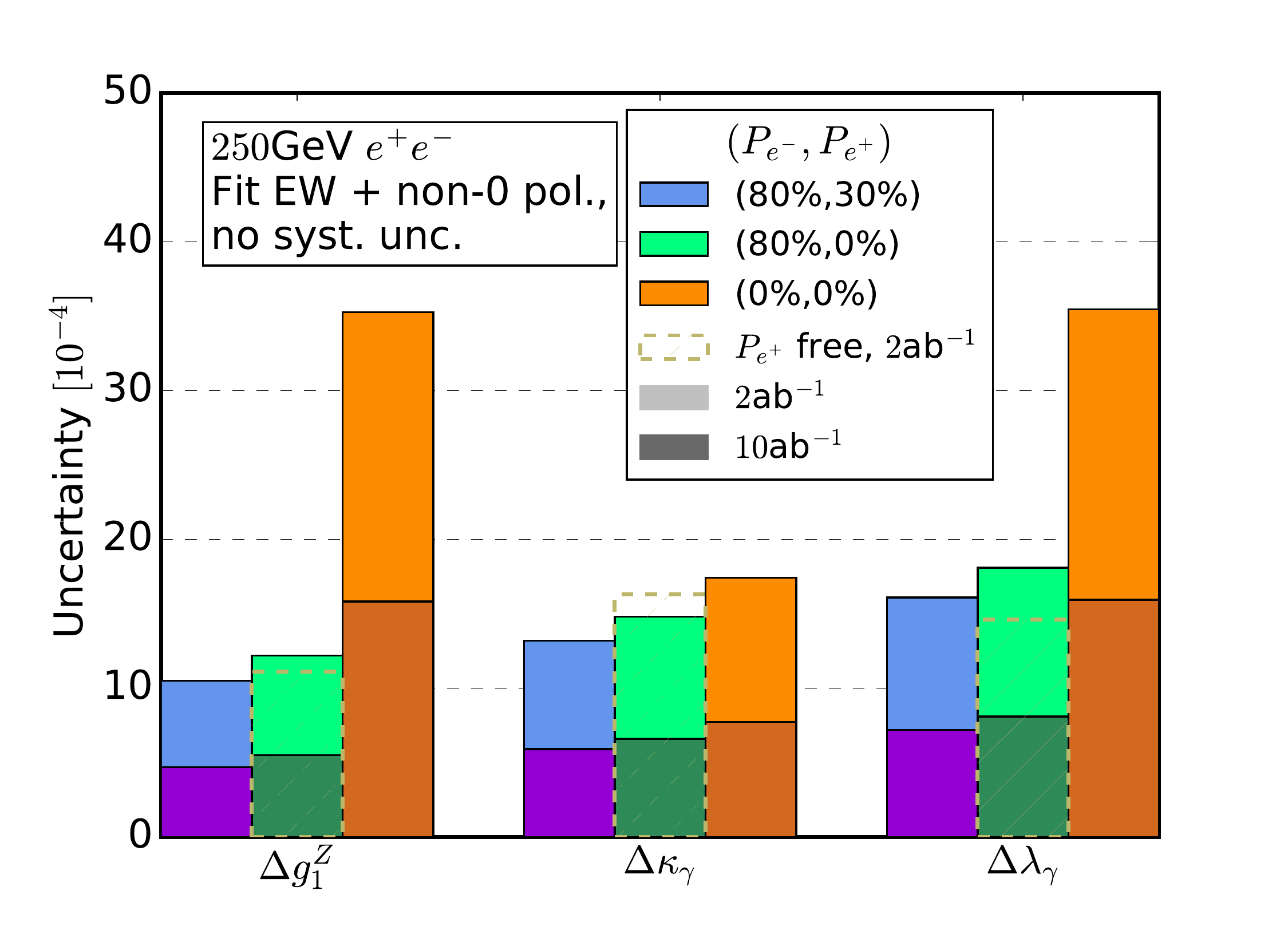}
    \caption{}
    \label{SUBFIG:CompTGC}
  \end{subfigure}%
  \begin{subfigure}[t]{0.5\textwidth}
    \centering
    \includegraphics[width=\textwidth]{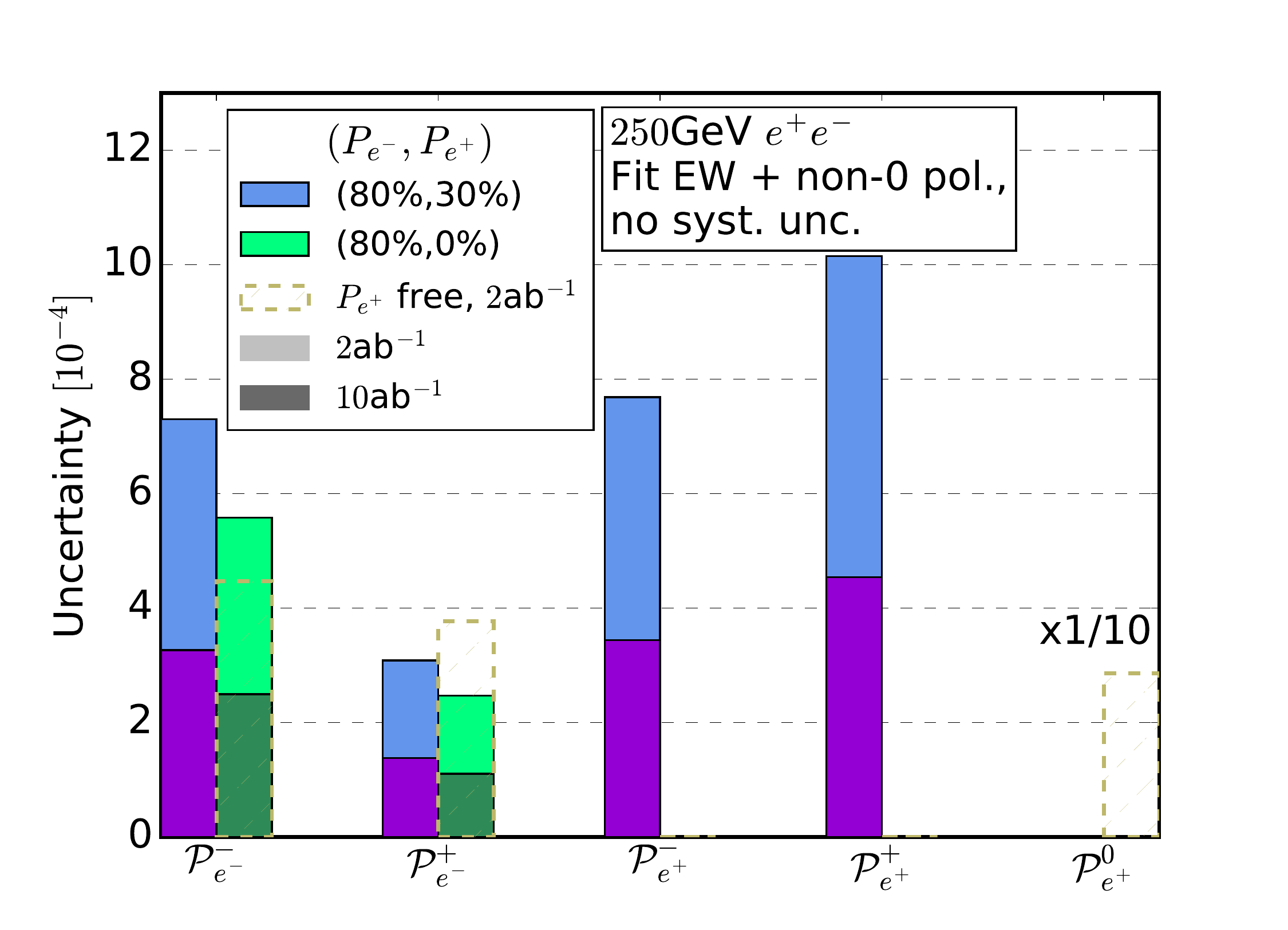}
    \caption{}
    \label{SUBFIG:CompPol}
  \end{subfigure}
  \begin{subfigure}[t]{0.5\textwidth}
    \centering
    \includegraphics[width=\textwidth]{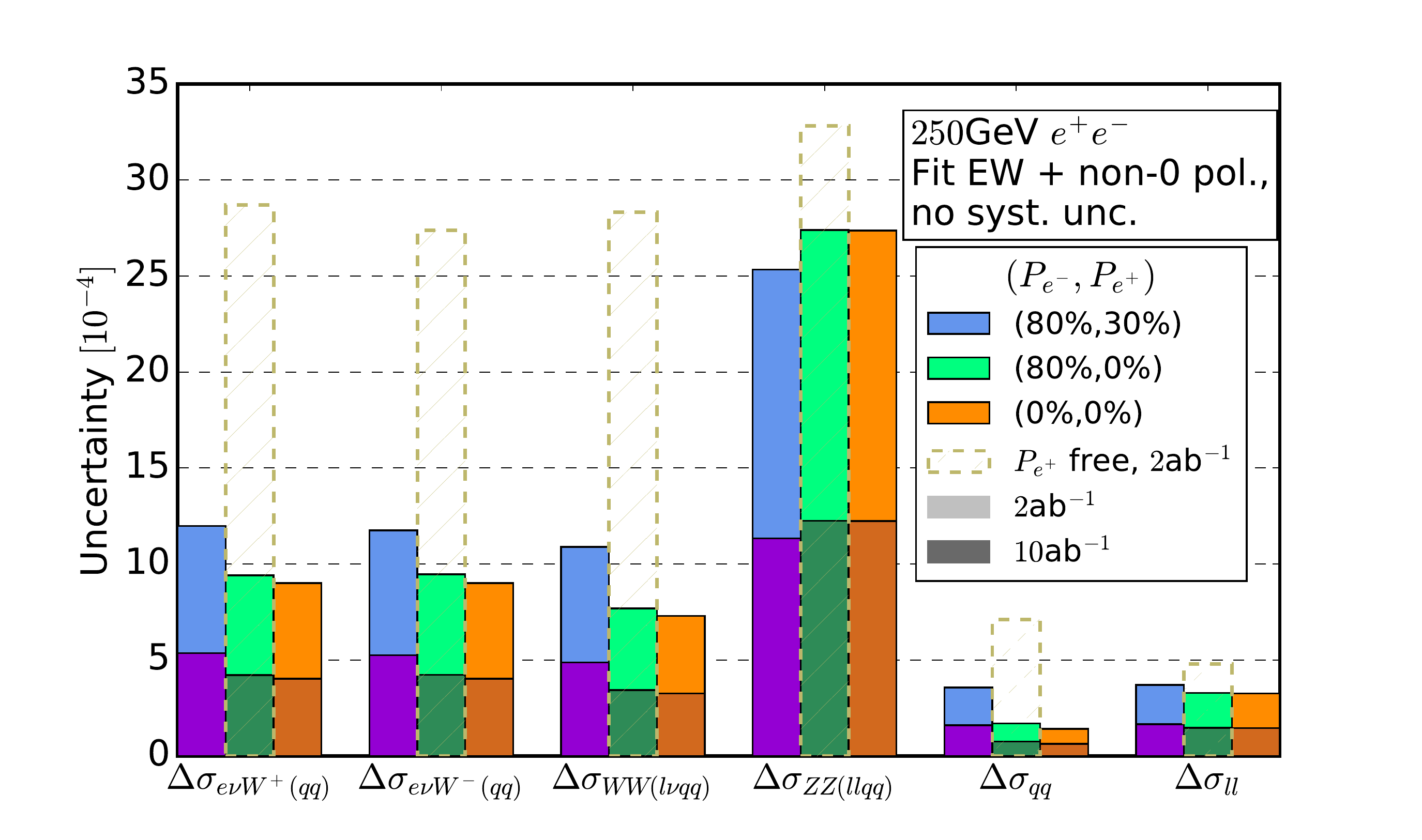}
    \caption{}
    \label{SUBFIG:CompAlpha}
  \end{subfigure}%
  \begin{subfigure}[t]{0.5\textwidth}
    \centering
    \includegraphics[width=\textwidth]{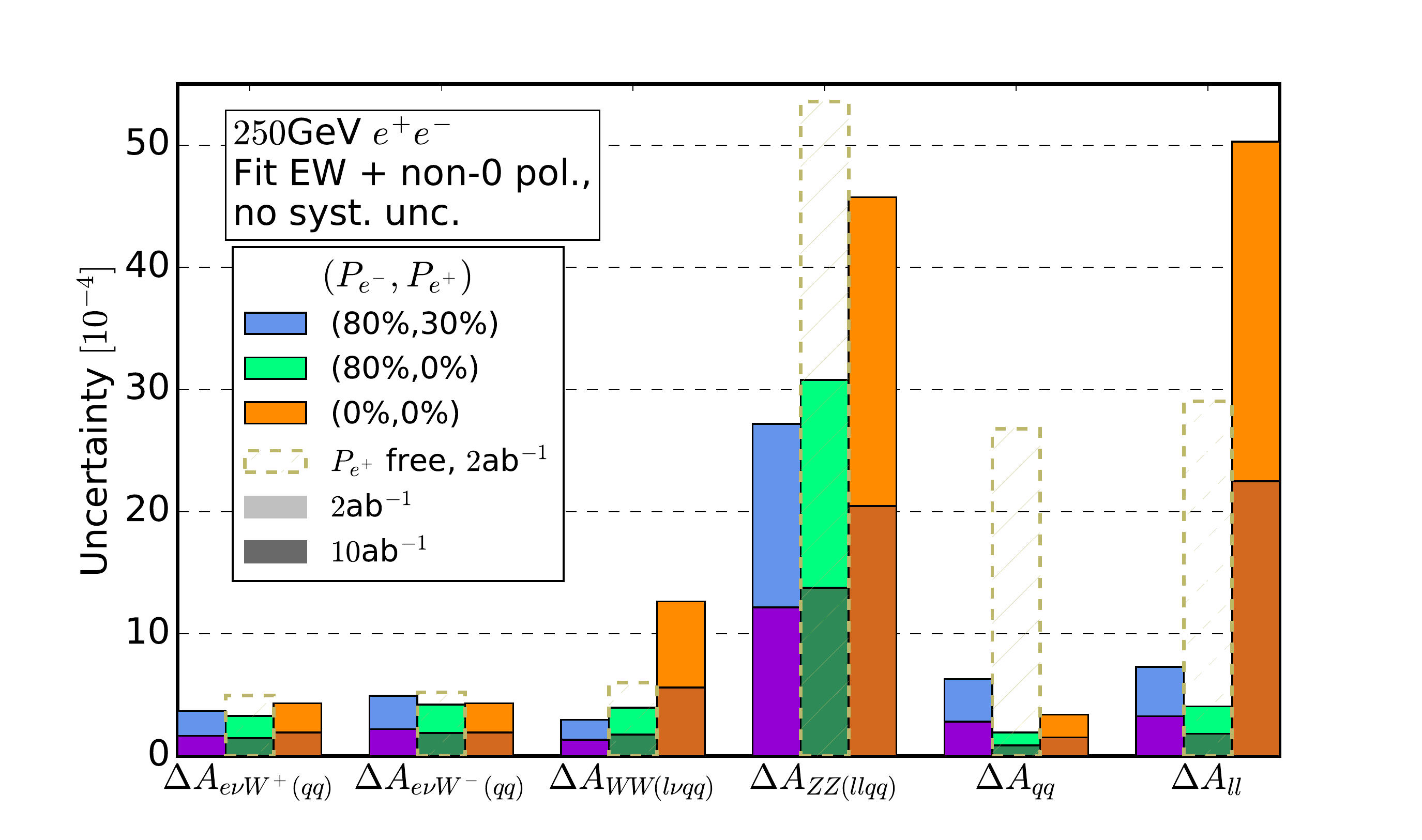}
    \caption{}
    \label{SUBFIG:CompBeta}
  \end{subfigure}
  \caption{
  Parameter uncertainties of \subfigref{SUBFIG:CompTGC} cTGC parameters, \subfigref{SUBFIG:CompPol} polarisations, \subfigref{SUBFIG:CompAlpha} chiral asymmetries and \subfigref{SUBFIG:CompBeta} sum of chiral cross sections for the different combinations of luminosity and available polarisations at $250\,$GeV.
  The uncertainties are averaged over 100 toy measurements.
  }
  \label{FIT:Comp}
\end{figure}

Here, fits are performed for a center-of-mass energy of $250\,$GeV.
Two integrated luminosities ($2\,$ab$^{-1}$, $10\,$ab$^{-1}$) and three polarisation cases (tab.~\ref{tab:lumisharing}) are studied.
The polarisation amplitude of unpolarised beams is by default fixed to zero and is not fitted.

One hundred toy measurements are produced for each collider setup to suppress possible statistical fluctuations from the toy measurement method.
A fit is performed on each toy measurement and the standard deviations for each fit parameter are extracted.
The resulting uncertainties are then calculated as the average standard deviations of all toy measurements (fig.~\ref{FIT:Comp}).
 
Variations of uncertainties between the toy measurements are observed to be at or below the percent level relative to the given uncertainty.
Any relative difference between uncertainties which is larger than a few percent is likely a systematic difference.

If no electron polarisation is available the $g_1^{Z}$ and $\lambda_{\gamma}$ uncertainties increase by more than a factor two for a fixed luminosity.
This can not be fully compensated by increasing the luminosity by a factor 5.
In the case of $\kappa_{\gamma}$, the unpolarised case has uncertainties less than a factor 1.5 worse than the polarised case.
Increasing the available statistics by increasing the luminosity is a more significant effect for this parameter.
The positron polarisation does not show a strong influence on the cTGC uncertainties in this framework.
Systematic uncertainties are expected to be affected if the positron beam is unpolarised.

Chiral cross section uncertainties improve by approximately a factor two when increasing the luminosity in every polarisation configuration (fig.~\ref{SUBFIG:CompAlpha},\ref{SUBFIG:CompBeta}).
This corresponds directly to the decreased statistical uncertainty.

Removing polarisations has a direct consequence for the extraction of asymmetries.
Having polarised beams means that asymmetries result in a change of cross section when flipping the sign of a polarisation.
An asymmetry can therefore be extracted by a simple total cross section measurement at different polarisations.
If no or fewer polarisations are available the asymmetry extraction relies on the measurement of the chiral composition of differential cross sections.
This may result in a loss of sensitivity.

Most asymmetry parameters show that sensitivity either stays in the same range or decreases by up to an order of magnitude if polarisations are removed.
A strong influence is seen in the asymmetries of the di-lepton production $A_{ll}$ and the semi-leptonic $Z$ boson pair decay $A_{l\nu qq}$.
This may be at least partially caused by the current artificial muon charge-symmetrisation of the angular distributions of the $qq'\mu\nu_{\mu}$ final state.
Separating the $\muP$ and $\muM$ distributions could soften this decrease of sensitivity.

In contrast, the total cross sections tend to show lower sensitivities when more polarisation is available.
This effect may be caused by the fixed zero-polarisations which are not measured in the fit.

The influence of measuring a close-to-zero beam polarisation in this scheme is tested for the single-polarisation case.
No effect on the uncertainty on the extracted cTGCs is observed.
The uncertainties on the chiral cross section parameters however become up to an order of magnitude higher than for the fully polarised case.

Polarisations are extracted with a sensitivity in the order of $\Delta\Pol/\Pol \sim 10^{-3} - 10^{-4}$ (fig.~\ref{SUBFIG:CompPol}). 
The measurement of a positron polarisation close to zero shows an uncertainty which is a factor $3-10$ higher than all others.
A polarimeter constraint may improve this situation~\cite{Fujii:2018mli}.

\section{Conclusion \& Outlook}

The fit demonstrates that the cTGC sensitivity is highly dependent on available luminosities and polarisations.
An absence of beam polarisation may complicate the measurement of cTGCs significantly and should be part of further studies.

This has so far been intended as a proof of principle to show that such comparisons can be made within a common fit framework.
The results described here do not yet include systematic uncertainties from analyses or detector effects.
Such systematic uncertainties may lead to significantly different changes for cases of polarised and unpolarised beams.
The splitting of unpolarised dataset into datasets with different initial polarisations usually results in a signal-enhanced and a signal-suppressed dataset.
Systematic effects can be investigated with the signal-suppressed dataset.
These effects can then be removed from the signal-enhanced dataset and the signal can be extracted.
How well this approach can be used to suppress systematic uncertainties remains an open field of study.

Further realism will be added to the current framework in order to better inform the design of future $\eP\eM$ colliders.

\section{Acknowledgments}

This work was funded by the Deutsche Forschungsgemeinschaft under Germany’s Excellence Strategy – EXC 2121 ``Quantum Universe'' – 390833306.

\section{References}
\customprintbibliography

\end{document}